\documentclass[letterpaper,11pt]{article}
\usepackage{fullpage}
\usepackage[english]{babel}
\usepackage{graphicx}
\usepackage{amssymb}
\usepackage{amsmath}
\usepackage{xspace}
\usepackage{wrapfig}

\graphicspath{{Figures/}}

\newenvironment{proofof}[1]{\emph{Proof of #1.}}{\hfill$\square$}

\newcommand{\remove}[1]{{}}

\newcommand{\length}[1]{\ensuremath{|#1|}}
\newcommand{\fullgraph}{full-\ensuremath{\theta_6}-graph\xspace}

\newcommand{\graph}{half-\ensuremath{\theta_6}-graph\xspace}
\newcommand{\canon}[2]{\ensuremath{T_{#1 #2}}}
\newcommand{\etal}{\emph{et al.}\xspace}

\newcommand{\hts}{\graph}

\newcommand{\dtw}{\ensuremath{G_{12}}\xspace}
\newcommand{\dn}{\ensuremath{G_{9}}\xspace}

\renewcommand{\c}[1]{\ensuremath{C_#1}}
\newcommand{\nc}[1]{\ensuremath{\overline{C}_#1}}

\newcommand{\sA}{\ensuremath{\mathcal{A}}\xspace}
\newcommand{\sB}{\ensuremath{\mathcal{B}}\xspace}
\newcommand{\sC}{\ensuremath{\mathcal{C}}\xspace}

\newtheorem{defin}{Definition}
  
\newtheorem{theo}[defin]{Theorem}
  \newenvironment{theorem}{\begin{theo} \sl}{\end{theo}}
\newtheorem{lem}[defin]{Lemma}
  \newenvironment{lemma}{\begin{lem} \sl}{\end{lem}}
\newtheorem{coro}[defin]{Corollary}
  \newenvironment{corollary}{\begin{coro} \sl}{\end{coro}}
\newtheorem{obs}[defin]{Observation}
  \newenvironment{observation}{\begin{obs} \sl}{\end{obs}}
\newenvironment{myproof}{\emph{Proof.}}{\hfill $\Box$ \medskip\\}

\newenvironment{shortitemize}
    {\begin{itemize}\setlength{\itemsep}{-5pt}}
    {\end{itemize}}

\newenvironment{shortenumerate}
    {\begin{enumerate}\setlength{\itemsep}{-5pt}}
    {\end{enumerate}}

{\makeatletter
 \gdef\and{\qquad}
 \gdef\maketitle{{
   \def\thefootnote{\@fnsymbol\c@footnote}
   \begin{center}
     \LARGE \@title
   \end{center}\medskip
   \centerline{\@author}\bigskip
   \@thanks
   \setcounter{footnote}{0}}}}

\author{Prosenjit Bose\thanks{School of Computer Science, Carleton University. Research supported in part by NSERC and Carleton University's President's 2010 Doctoral Fellowship. Email: \texttt{jit@scs.carleton.ca, andre@cg.scs.carleton.ca, sander@cg.scs.carleton.ca}. }
\and
Rolf Fagerberg\thanks{Department of Mathematics and Computer Science, University of Southern Denmark. Email: \texttt{rolf@imada.sdu.dk}. Partially supported by the Danish Council for Independent Research, Natural Sciences.}
\and 
\addtocounter{footnote}{-2}
Andr\'e van Renssen\footnotemark
\and
\addtocounter{footnote}{-1}
Sander Verdonschot\footnotemark
}
\title{Optimal local routing on Delaunay triangulations defined by empty equilateral triangles}
\date{\today}

\begin{document}

\maketitle

\begin{abstract}
 We present a deterministic local routing algorithm that is guaranteed to find a path between any pair of vertices in a half-$\theta_6$-graph\footnote{Extended abstracts containing the results in this paper appeared in SODA 2012 and CCCG 2012.} (the half-$\theta_6$-graph is equivalent to the Delaunay triangulation where the empty region is an equilateral triangle). The length of the path is at most $5/\sqrt{3} \approx 2.887$ times the Euclidean distance between the pair of vertices. Moreover, we show that no local routing algorithm can achieve a better routing ratio, thereby proving that our routing algorithm is optimal. This is somewhat surprising because the spanning ratio of the half-$\theta_6$-graph is 2, meaning that even though there always exists a path whose lengths is at most twice the Euclidean distance, we cannot always find such a path when routing locally.
 
 Since every triangulation can be embedded in the plane as a half-$\theta_6$-graph using $O(\log n)$ bits per vertex coordinate via Schnyder's embedding scheme (SODA 1990), our result provides a competitive local routing algorithm for every such embedded triangulation. Finally, we show how our routing algorithm can be adapted to provide a routing ratio of $15/\sqrt{3} \approx 8.660$ on two bounded degree subgraphs of the half-$\theta_6$-graph.
\end{abstract}

\thispagestyle{empty}

\newpage

\setcounter{page}{1}

\section{Introduction}

A fundamental problem in networking is the routing of a message from one vertex to another in a graph. What makes routing more challenging is that often in a network the routing strategy must be \emph{local}. Informally, a routing strategy is \emph{local} when the routing algorithm must choose the next vertex to forward a message to based solely on knowledge of the source and destination vertex, the current vertex and all vertices directly connected to the current vertex. Routing algorithms are considered \emph{geometric} when the graph is embedded in the plane, with edges being straight line segments connecting pairs of points and weighted by the Euclidean distance between their endpoints. Geometric routing algorithms are important in wireless sensor networks \mbox{(see \cite{G09} and \cite{R09}} for surveys of the area) since they offer routing strategies that use the coordinates of the vertices to help guide the search as opposed to using the more traditional routing tables.  

Papadimitriou and Ratajczak~\cite{PR05} posed a tantalizing question in this area that lead to a flurry of activity: Does every 3-connected planar graph have a straight-line embedding in the plane that admits a local routing strategy such as greedy routing\footnote{A routing strategy is greedy when a message is always forwarded to the vertex whose distance to the destination is the smallest among all vertices in the neighborhood of the current vertex, including the current vertex.}? They provided a partial answer by showing that 3-connected planar graphs can always be embedded in $\mathbb{R}^3$ such that they admit a greedy routing strategy. They also showed that the class of complete bipartite graphs, $K_{k,6k+1}$ for all $k\geq 1$ cannot be embedded such that greedy routing always succeeds since every embedding has at least one vertex that is not connected to its nearest neighbor. Bose and Morin~\cite{BM04} showed that greedy routing always succeeds on Delaunay triangulations. In fact, a slightly restricted greedy routing strategy known as {\em greedy-compass} is the first local routing strategy shown to succeed on all triangulations~\cite{Betal02}. Dhandapani~\cite{D10} proved the existence of an embedding that admits greedy routing for every triangulation and Angelini~\etal~\cite{AFG10} provided a constructive proof. Leighton and Moitra~\cite{LM10} settled Papadimitriou and Ratajczak's question by showing that every 3-connected planar graph can be embedded in the plane such that greedy routing succeeds. One drawback of these embedding algorithms is that the coordinates require $\Omega(n \log n)$ bits per vertex. To address this, He and Zhang~\cite{HZ10} and Goodrich and Strash~\cite{GS08} gave succinct embeddings using only $O(\log n)$ bits per vertex. Recently, He and Zhang~\cite{HZ11} showed that every 3-connected plane graph admits a succinct embedding with convex faces on which a slightly modified greedy routing strategy always succeeds.

In light of these recent successes, it is surprising to note that the above routing strategies have solely concentrated on finding an embedding that guarantees that a local routing strategy will succeed, but pay little attention to the quality of the resulting path. For example, none of the above routing strategies have been shown to be \emph{competitive}. A geometric routing strategy is said to be competitive if the length of the path found by the routing strategy is not more than a constant times the Euclidean distance between its endpoints. This constant is called the \emph{routing ratio}. Bose and Morin~\cite{BM04} show that many local routing strategies are not competitive, but show how to route competitively on the Delaunay triangulation. However, Dillencourt~\cite{D90} showed that not all triangulations can be embedded in the plane as Delaunay triangulations. This raises the following question: can \textbf{every} triangulation be embedded in the plane such that it admits a competitive local routing strategy? We answer this question in the affirmative.

The \graph was introduced by Bonichon~\etal~\cite{BGHI10}, who showed that it is identical to the Delaunay triangulation where the empty region is an equilateral triangle. Although both graphs are identical, the local definition of the \graph makes it more useful in the context of routing. We formally the \graph in the next section. Our main result is a deterministic local routing algorithm that is guaranteed to find a path between any pair of vertices in a \graph whose length is at most $5/\sqrt{3}$ times the Euclidean distance between the pair of vertices. On the way to proving our main result, we uncover some local properties of spanning paths in the \graph. Since Schnyder~\cite{S90} showed that every triangulation can be embedded in the plane as a \graph using $O(\log n)$ bits per vertex coordinate, our main result implies that every triangulation has an embedding that admits a competitive local routing algorithm. Moreover, we show that no local routing algorithm can achieve a better routing ratio on a \graph, implying that our routing algorithm is optimal. This is somewhat surprising because Chew~\cite{Chew89} showed that the spanning ratio of the \graph is 2. Thus, our lower bound provides a separation between the spanning ratio of the \graph and the best achievable routing ratio on the \graph. We believe that this is the first separation between the spanning ratio and routing ratio of any graph. It also makes the \hts one of the few graphs for which tight spanning and routing ratios are known. Finally, we show how our routing algorithm can be adapted to provide a routing ratio of $15/\sqrt{3}$ on two bounded degree subgraphs of the \graph introduced by Bonichon~\etal~\cite{BGHP10}. To the best of our knowledge, this is the first competitive routing algorithm on a bounded-degree plane graph.

\section{Preliminaries} \label{sec:prelims}

In order to find a competitive path between any two vertices of a graph, such a path must first exist. Graphs that meet this criterion are called \emph{spanners}. Formally, given a weighted graph $G$, we define the distance $d_G(u, v)$ between two vertices $u$ and $v$ to be the sum of the weights of the edges in the shortest path between $u$ and $v$ in $G$. A subgraph $H$ of $G$ is a \emph{$t$-spanner} of $G$ if for all pairs of vertices $u$ and $v$, $d_H(u, v) \leq t \cdot d_G(u, v)$, for $t\geq 1$. We say that $H$ is a \emph{spanner} if it is a $t$-spanner for some constant $t$. The \emph{spanning ratio} of $H$ is the smallest $t$ for which it is a $t$-spanner. The graph $G$ is referred to as the \emph{underlying graph}. 


Unless otherwise noted, we assume that the underlying graph $G$ is a straight-line embedding of the complete graph on a set of $n$ points in the plane, with the weight of an edge $(u,v)$ being the Euclidean distance~$|u v|$ between $u$ and $v$. A spanner of such a graph is called a \emph{geometric spanner}. We focus on one specific class of geometric spanners: the \hts. In a slight abuse of notation, we often speak about the spanning ratio of the \hts. By this, we mean the maximum spanning ratio of any \hts on any set of $n$ points in the plane. In the remainder of this section, we describe the construction of the \hts and introduce some notation.

\begin{figure}[ht]
 \centering
 \includegraphics{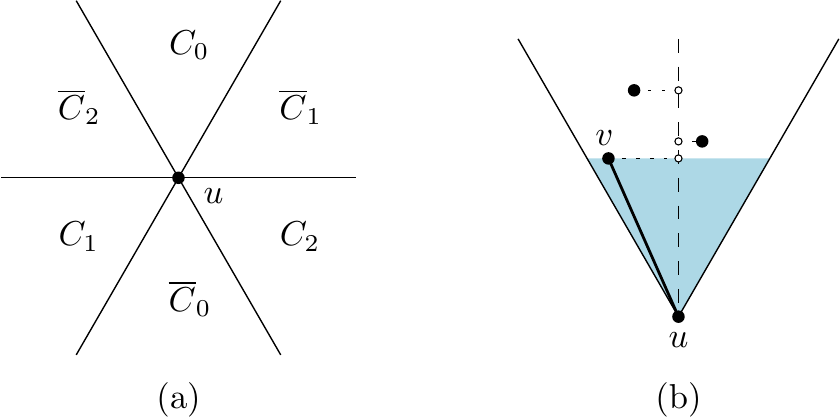}
 \caption{(a) The cones around a vertex $u$. (b) The construction of the \hts. In each positive cone, $u$ connects to the vertex with the closest projection on the bisector of that cone.}
 \label{fig:cones}
\end{figure}

Given a set $P$ of points in the plane, we consider each point $u \in P$ and partition the plane into 6 cones (regions in the plane between two rays originating from the same point) with apex $u$, each defined by two rays at consecutive multiples of $\pi/3$ radians from the positive $x$-axis. We label the cones $\nc{1}$, $\c{0}$, $\nc{2}$, $\c{1}$, $\nc{0}$ and $\c{2}$, in counter-clockwise order around $u$, starting from the positive $x$-axis \mbox{(see Figure~\ref{fig:cones}a)}. The cones $\c{0}$, $\c{1}$ and $\c{2}$ are called \emph{positive}, while the others are called \emph{negative}. When the apex is not clear from the context, we use $C^u_i$ to denote cone $C_i$ with apex $u$.

To build the \hts, we consider each vertex $u$ and add an edge to the `closest' vertex in each of its positive cones. However, instead of using the Euclidean distance, we measure distance by projecting each vertex in the cone onto the bisector of the cone. We call the vertex in this cone whose projection is closest to $u$ the \emph{closest vertex} and connect it to $u$ with an edge (see Figure~\ref{fig:cones}b). For simplicity, we assume that no two points lie on a line parallel to a cone boundary, guaranteeing that each vertex connects to exactly one vertex in each positive cone. Hence the graph has at most $3n$ edges in total.

Given two vertices $u$ and $v$ such that $v$ lies in a positive cone of $u$, we define their \emph{canonical triangle} $T_{uv}$ to be the triangle bounded by the cone of $u$ that contains $v$ and the line through $v$ perpendicular to the bisector of that cone. For example, the shaded region in Figure~\ref{fig:cones}b is the canonical triangle of $u$ and $v$. Note that for any pair of vertices $u$ and $v$, either $v$ lies in a positive cone of $u$, or $u$ lies in a positive cone of $v$, so there is exactly one canonical triangle (either \canon{u}{v} or \canon{v}{u}) for the pair. The construction of the \hts can alternatively be described as adding an edge between two vertices if and only if their canonical triangle is empty. This property will play an important role in our proofs.

\section{Spanning ratio of the \graph} \label{sec:spanning}

Bonichon~\etal~\cite{BGHI10} showed that the \graph is a geometric spanner with spanning ratio~2 by showing it is equivalent to the Delaunay triangulation based on empty equilateral triangles, which is known to have spanning ratio~2~\cite{Chew89}. This correspondence also shows that the \graph is internally triangulated: every face except for the outer face is a triangle (this follows from the duality with the Voronoi diagram, along with the fact that all vertices in the Voronoi diagram have degree 3, provided that no 4 points lie on the same equilateral triangle). In this section, we provide an alternative proof of the spanning ratio of the \graph. Our proof shows that between any pair of points, there always exists a path with spanning ratio 2 that lies in the canonical triangle. This property plays an important role in our routing algorithm, which we describe in Section~\ref{sec:routing}.

\begin{figure}[ht]
    \begin{center}
      \includegraphics{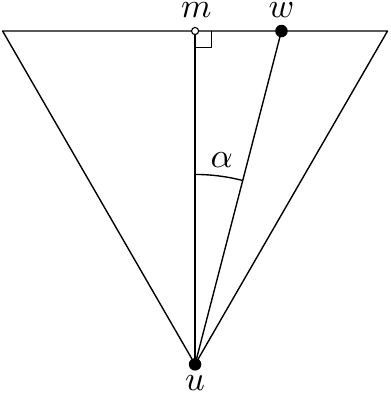}
    \end{center}
    \caption{Two vertices $u$ and $w$ with their canonical triangle $\canon{u}{w}$. The angle $\alpha$ is the unsigned angle between the line $uw$ and the bisector of the cone containing $w$.}
    \label{fig:angleFigure}
  \end{figure}

For a pair of vertices~$u$ and $w$, our bound is expressed in terms of the angle~$\alpha$ between the line from $u$ to $w$ and the bisector of their canonical triangle (see~Figure~\ref{fig:angleFigure}).

\begin{theo}
\label{theo:UnconstrainedSpanningRatio}
Let $u$ and $w$ be vertices with $w$ in a positive cone of $u$. Let $m$ be the midpoint of the side of \canon{u}{w} opposing $u$, and let $\alpha$ be the unsigned angle between $uw$ and $um$. There exists a path between $u$ and $w$ in the \graph, of length at most
$$(\sqrt{3} \cdot \cos \alpha + \sin \alpha) \cdot |u w|,$$
where all vertices on this path lie in $\canon{u}{w}$.
\end{theo}
The expression $\sqrt{3} \cdot \cos \alpha + \sin \alpha$ is increasing for $\alpha \in [0,\pi/6]$. Inserting the extreme value $\pi/6$ for $\alpha$, we arrive at the following.
\begin{coro}
\label{cor:spanning}
The spanning ratio of the \graph is 2.
\end{coro}
We note that the bounds of Theorem~\ref{theo:UnconstrainedSpanningRatio} and Corollary~\ref{cor:spanning} are tight: for all values of $\alpha \in [0,\pi/6]$ there exists a point set for which the shortest path in the \graph for some pair of vertices $u$ and $w$ has length arbitrarily close to $(\sqrt{3} \cdot \cos \alpha + \sin \alpha) \cdot |u w|$. A simple example appears later in the proof of Theorem~\ref{thm:routing}.

\vspace{\medskipamount}
\begin{proofof}{Theorem~\ref{theo:UnconstrainedSpanningRatio}}
  Given two vertices $u$ and $w$, we assume without loss of generality that $w$ lies in $C^u_0$. We prove the theorem by induction on the rank, when ordered by area, of the triangles $\canon{x}{y}$ for all pairs of points $x$ and $y$ where $y$ lies in a positive cone of $x$. Let $a$ and $b$ be the upper left and right corner of $\canon{u}{w}$, and let $A = \canon{u}{w} \cap C^w_1$ and $B = \canon{u}{w} \cap C^w_2$, as illustrated in Figure~\ref{fig:Triangle}.

  \begin{figure}[ht]
    \begin{center}
      \includegraphics{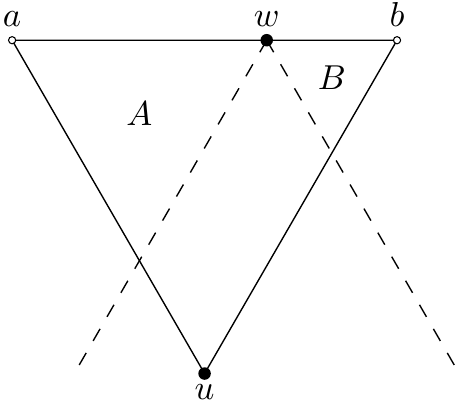}
    \end{center}
    \caption{The corners $a$ and $b$, and the regions $A$ and $B$.}
    \label{fig:Triangle}
  \end{figure}

  Our inductive hypothesis is the following, where $\delta(u,w)$ denotes the length of the shortest path from~$u$ to $w$ in the part of the \graph induced by the vertices in \canon{u}{w}.
  \begin{shortenumerate}
    \item If $A$ is empty, then $\delta(u, w) \leq |u b| + |b w|$.
    \item If $B$ is empty, then $\delta(u, w) \leq |u a| + |a w|$.
    \item If neither $A$ nor $B$ is empty, then $\delta(u, w) \leq \max\{|u a| + |a w|, |u b| + |b w|\}$.
  \end{shortenumerate}

  We first note that this induction hypothesis implies Theorem~\ref{theo:UnconstrainedSpanningRatio}: using the side of  \canon{u}{w} as the unit of length, we have from Figure~\ref{fig:angleFigure} that $\length{wm} = \length{uw}\cdot\sin\alpha$ and $\sqrt{3}/2 = \length{um} = \length{uw}\cdot\cos\alpha$. Hence the induction hypothesis gives us that $\delta(u, w)$ is at most $1+1/2+\length{wm} = \sqrt{3} \cdot (\sqrt{3}/2) +\length{wm} = (\sqrt{3} \cdot \cos\alpha + \sin\alpha) \cdot \length{uw}$, as required.

  \textbf{Base case:} $\canon{u}{w}$ has rank 1. Since there are no smaller canonical triangles, $w$ must be the closest vertex to $u$. Hence the edge $(u,w)$ is in the \graph, and $\delta(u, w) = |u w|$. Using the triangle inequality, we have $|u w| \leq \min\{|u a| + |a w|, |u b| + |b w|\}$, so the induction hypothesis holds.

  \textbf{Induction step:} We assume that the induction hypothesis holds for all pairs of points with canonical triangles of rank up to $i$. Let $\canon{u}{w}$ be a canonical triangle of rank $i+1$.

  If $(u,w)$ is an edge in the \graph, the induction hypothesis follows by the same argument as in the base case. If there is no edge between $u$ and $w$, let $v$ be the vertex closest to $u$ in the positive cone $C^u_0$, and let $a'$ and $b'$ be the upper left and right corner of $\canon{u}{v}$. By definition, $\delta(u, w) \leq |u v| + \delta(v, w)$, and by the triangle inequality, $|u v| \leq \min\{|u a'| + |a' v|, |u b'| + |b' v|\}$.

  We perform a case distinction on the location of $v$: (a) $v$ lies  neither in $A$ nor in $B$, (b) $v$ lies inside $A$, and (c) $v$ lies inside $B$. The case where $v$ lies inside $B$ is analogous to the case where $v$ lies inside $A$, so we only discuss the first two cases, which are illustrated in Figure~\ref{fig:TriangleCases2}.

  \begin{figure}[ht]
    \begin{center}
      \includegraphics{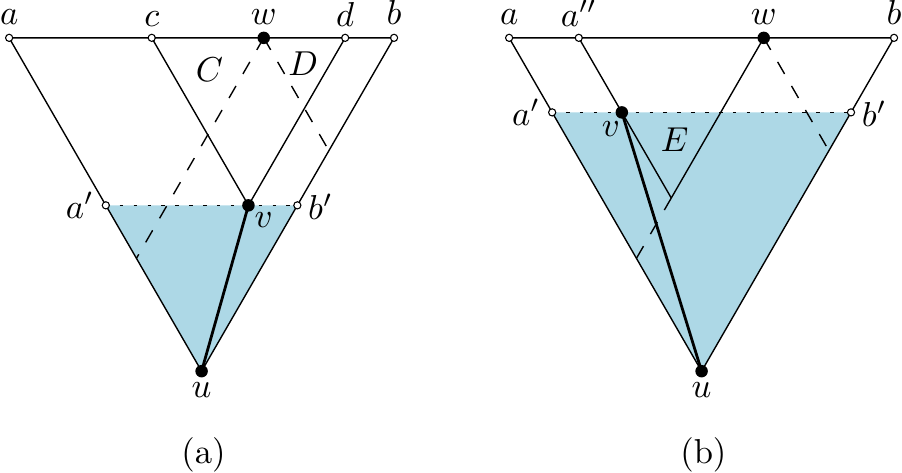}
    \end{center}
    \caption{The two cases: (a) $v$ lies in neither $A$ nor $B$, (b) $v$ lies in $A$.}
    \label{fig:TriangleCases2}
  \end{figure}

 \textbf{Case (a):} Let $c$ and $d$ be the upper left and right corner of $\canon{v}{w}$, and let $C = \canon{v}{w} \cap C^w_1$ and $D = \canon{v}{w} \cap C^w_2$ (see~Figure~\ref{fig:TriangleCases2}a). Since $\canon{v}{w}$ has smaller area than $\canon{u}{w}$, we apply the inductive hypothesis on $\canon{v}{w}$. Our task is to prove all three statements of the inductive hypothesis for $\canon{u}{w}$.

  \begin{enumerate}
  \item If $A$ is empty, then $C$ is also empty, so by induction $\delta(v, w) \leq |v d| + |d w|$. Since $v$, $d$, $b$, and $b'$ form a parallelogram, we have:
   \begin{eqnarray*}
    \delta(u, w) &\leq& |u v| + \delta(v, w) \\
    &\leq& |u b'| + |b' v| + |v d| + |d w| \\
    &=& |u b| + |b w|,
   \end{eqnarray*}
  which proves the first statement of the induction hypothesis. This argument is illustrated in Figure~\ref{fig:SpanningProofVisualization}a.

      \item If $B$ is empty, an analogous argument proves the second statement of the induction hypothesis.

      \item If neither $A$ nor $B$ is empty, by induction we have $\delta(v, w) \leq \max\{|v c| + |c w|, |v d| + |d w|\}$. Assume, without loss of generality, that the maximum of the right hand side is attained by its second argument $|v d| + |d w|$ (the other case is analogous).

      Since vertices $v$, $d$, $b$, and $b'$ form a parallelogram, we have that:
      \begin{eqnarray*}
	\delta(u, w) &\leq& |uv| + \delta(v, w) \\
	&\leq& |u b'| + |b' v| +  |v d| + |d w| \\
	&\leq& |u b| + |b w| \\
	&\leq& \max\{|u a| + |a w|, |u b| + |b w|\},
      \end{eqnarray*}
      which proves the third statement of the induction hypothesis. This argument is illustrated in Figure~\ref{fig:SpanningProofVisualization}b.
  \end{enumerate}

  \begin{figure}[ht]
    \begin{center}
      \includegraphics{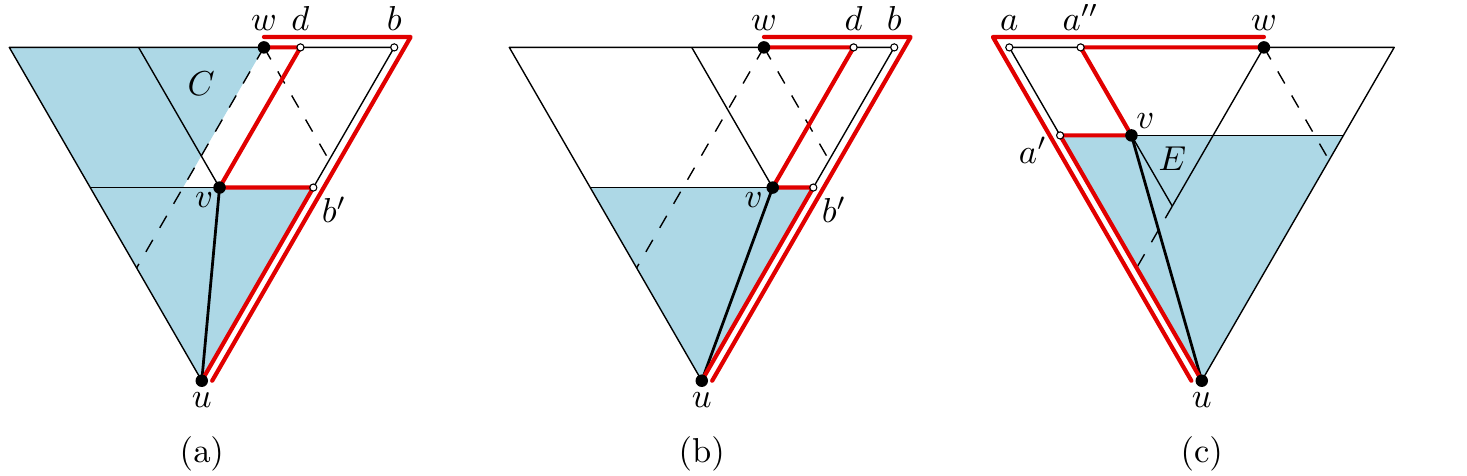}
    \end{center}
    \caption{Visualization of the path inequalities in three cases: (a) $v$ lies in neither $A$ nor $B$ and one of $A$ or $B$ is empty (cases a.1 and a.2 in our proof), (b) $v$ lies in neither $A$ nor $B$ and neither is empty (case a.3), (c) $v$ lies in $A$ or $B$ (case b). The paths occurring in the equations are drawn with thick red lines, and light blue areas indicate empty regions.}
    \label{fig:SpanningProofVisualization}
  \end{figure}

  \textbf{Case (b):} Let $E = \canon{u}{v} \,\cap\, \canon{w}{v}$, and let $a''$ be the upper left corner of $\canon{w}{v}$ (see~Figure~\ref{fig:TriangleCases2}b). Since $v$ is the closest vertex to $u$ in one of its positive cones, \canon{u}{v} is empty and hence $E$ is also empty. Since $\canon{w}{v}$ is smaller than $\canon{u}{w}$, we can apply induction on it. As $E$ is empty, the first statement of the induction hypothesis for $\canon{w}{v}$ applies, giving us that $\delta(v, w) \leq |v a''| + |a'' w|$. Since $|u v| \leq |u a'| + |a' v|$ and $v$, $a''$, $a$, and $a'$ form a parallelogram, we have that $\delta(u, w) \leq |u a| + |a w|$, proving the second and third statement in the induction hypothesis for \canon{u}{w}. This argument is illustrated in Figure~\ref{fig:SpanningProofVisualization}c. Since $v$ lies in $A$, the first statement in the induction hypothesis for \canon{u}{w} is vacuously true.
\end{proofof}

\section{Remarks on the spanning ratio} \label{sec:remarks}

The \fullgraph, introduced by Keil and Gutwin~\cite{KG92}, is similar to the \graph except that all 6 cones are positive cones. Thus, the \fullgraph is the union of two copies of the \graph, where one \graph is rotated by $\pi/3$ radians. The \graph and the \fullgraph both have a spanning ratio of 2, with lower bound examples showing that it is tight for both graphs. This is surprising since the \fullgraph can have twice the number of edges of the \graph.

Note that since the \fullgraph consists of two rotated copies of the \graph, one question that comes to mind is what is the best spanning ratio if one is to construct a graph consisting of two rotated copies of the \graph ? Can one do better than a spanning ratio of 2? Consider the following construction. Build two \graph{}s as described in Section~\ref{sec:prelims}, but rotate each cone of the second graph by $\pi/6$ radians. For each pair of vertices, there is a path of length at most $\sqrt{3}\cos{\alpha} + \sin{\alpha}$ times the Euclidean distance between them, where $\alpha$ is the angle between the line connecting the vertices in question, and the closest bisector. Since this function is increasing, the spanning ratio is defined by the maximum possible angle to the closest bisector, which is $\pi/12$ radians, giving a spanning ratio of roughly 1.932.

By using $k$ copies, we improve the spanning ratio even further: if each is rotated by $\pi/(3k)$ radians, we get a spanning ratio of $\sqrt{3}\cos{\frac{\pi}{6k}} + \sin{\frac{\pi}{6k}}$. This is better than the known upper bounds for the full $\theta_{3k}$-graph~\cite{bose2013spanning} for $k \leq 3$ and for the Yao$_{3k}$-graph~\cite{barba2014new} for $k \leq 4$.

\begin{corollary}
 The union of $k$ copies of the \graph{}, each rotated by $\pi/(3k)$ radians, is a geometric spanner with up to $3k$ edges and spanning ratio at most $\sqrt{3}\cos{\frac{\pi}{6k}} + \sin{\frac{\pi}{6k}}$.
\end{corollary}

\section{Routing in the \graph} \label{sec:routing}

In this section, we give matching upper and lower bounds for the routing ratio on the \graph. We begin by defining our model. Formally, a routing algorithm $A$ is a deterministic $k$-local, $m$-memory routing algorithm, if the vertex to which a message is forwarded from the current vertex $s$ is a function of $s$, $t$, $N_k(s)$, and $M$, where $t$ is the destination vertex, $N_k(s)$ is the $k$-neighborhood of $s$ and $M$ is a memory of size $m$, stored with the message. The $k$-neighborhood of a vertex $s$ is the set of vertices in the graph that can be reached from $s$ by following at most $k$ edges. For our purposes, we consider a unit of memory to consist of a $\log_2 n$ bit integer or a point in $\mathbb{R}^2$. Our model also assumes that the only information stored at each vertex of the graph is $N_k(s)$. Since our graphs are geometric, we identify each vertex by its coordinates in the plane. A routing algorithm is {\em $d$-competitive} provided that the total distance travelled by the message is never more than $d$ times the Euclidean distance between source and destination. Analogous to the spanning ratio, the \emph{routing ratio} of an algorithm is the smallest $d$ for which it is $d$-competitive.

We present a deterministic $1$-local $0$-memory algorithm that achieves the upper bounds, but our lower bounds hold for any deterministic $k$-local $0$-memory algorithm. Our bounds are expressed in terms of the angle~$\alpha$ between the line from the source to the destination and the bisector of their canonical triangle (see~Figure~\ref{fig:angleFigure}).

\begin{theo}\label{thm:routing}
Let $u$ and $w$ be two vertices, with $w$ in a positive cone of $u$. Let $m$ be the midpoint of the side of \canon{u}{w} opposing $u$, and let $\alpha$ be the unsigned angle between $uw$ and $um$. There is a deterministic $1$-local $0$-memory routing algorithm on the \graph for which every path followed has length at most
\renewcommand{\labelenumi}{{\rm\roman{enumi})}}
\begin{enumerate}
\label{thm:routing:item:posbound}
\item $(\sqrt{3} \cdot \cos \alpha + \sin \alpha) \cdot |u w|$ when routing from $u$ to $w$,
\label{thm:routing:item:negbound}
\item $(5/\sqrt{3} \cdot \cos \alpha - \sin \alpha) \cdot |u w|$ when routing from $w$ to $u$,
\end{enumerate}
and this is best possible for deterministic $k$-local, $0$-memory routing algorithms, where $k$ is constant.
\end{theo}
The first expression is increasing for $\alpha \in [0,\pi/6]$, while the second expression is decreasing. Inserting the extreme values $\pi/6$ and $0$ for $\alpha$, we get the following worst case version of Theorem~\ref{thm:routing}.
\begin{coro}
Let $u$ and $w$ be two vertices, with $w$ in a positive cone of $u$. There is a deterministic $1$-local $0$-memory routing algorithm on the \graph with routing ratio
\renewcommand{\labelenumi}{{\rm\roman{enumi})}}
\begin{enumerate}
\item $2$ when routing from $u$ to $w$,
\item $5/\sqrt{3} = 2.886\dots$ when routing from $w$ to $u$,
\end{enumerate}
and this is best possible for deterministic $k$-local, $0$-memory routing algorithms, where $k$ is constant.
\end{coro}
Since the spanning ratio of the \graph is~2, the second lower bound shows a separation between the spanning ratio and the best possible routing ratio in the \graph.

Since every triangulation can be embedded in the plane as a half-$\theta_6$-graph using $O(\log n)$ bits per vertex via Schnyder's embedding scheme~\cite{S90}, an important implication of Theorem~\ref{thm:routing} is the following.

\begin{coro}
Every $n$-vertex triangulation can be embedded in the plane using $O(\log n)$ bits per coordinate such that the embedded triangulation admits a deterministic $1$-local $0$-memory routing algorithm with routing ratio at most $5/\sqrt{3}$.
\end{coro}

In the remainder of this section we prove Theorem~\ref{thm:routing}. We split the proof into two cases, depending on whether the destination lies in a positive (Section~\ref{sec:positiveRouting}) or negative (Section~\ref{sec:negativeRouting}) cone of the source. In each case, we first present a proof of the lower bound, then a description of the routing algorithm, and finally a proof of the upper bound.

\subsection{Positive routing}
\label{sec:positiveRouting}

\begin{figure}[ht]
  \begin{center}
   \includegraphics{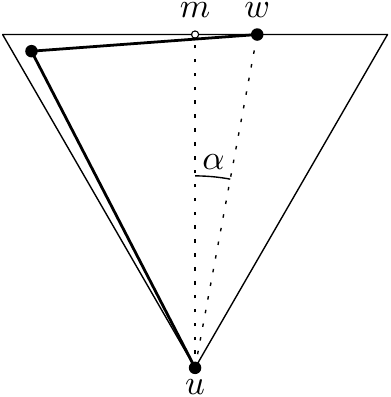}
  \end{center}
  \caption{The lower bound example when routing to a vertex in a positive cone.}
  \label{fig:lowerBoundInstancesPos}
\end{figure}

\begin{lemma}\textbf{\emph{(Lower bound for positive routing)}} \label{lem:poslb}
 Let $u$ and $w$ be two vertices, with $w$ in a positive cone of $u$. Let $m$ be the midpoint of the side of \canon{u}{w} opposing $u$, and let $\alpha$ be the unsigned angle between $uw$ and $um$. For any routing algorithm, there are instances for which the path followed has length at least $(\sqrt{3} \cdot \cos \alpha + \sin \alpha) \cdot |u w|$ when routing from $u$ to $w$.
\end{lemma}
\begin{myproof}
 Let the side of \canon{u}{w} be the unit of length. From Figure~\ref{fig:angleFigure}, we have $\length{wm} = \length{uw}\cdot\sin\alpha$ and $\sqrt{3}/2 = \length{um} = \length{uw}\cdot\cos\alpha$. From Figure~\ref{fig:lowerBoundInstancesPos}, the spanning ratio of the \graph is at least $1+1/2+\length{wm} = \sqrt{3} \cdot (\sqrt{3}/2) +\length{wm} = (\sqrt{3} \cdot \cos\alpha + \sin\alpha) \cdot \length{uw}$, since the point in the upper left corner of~\canon{u}{w} can be moved arbitrarily close to the corner. As there is no shorter path between $u$ and $w$, this is a lower bound for \emph{any} routing algorithm.
\end{myproof}

\begin{figure}[ht]
  \begin{center}
   \includegraphics{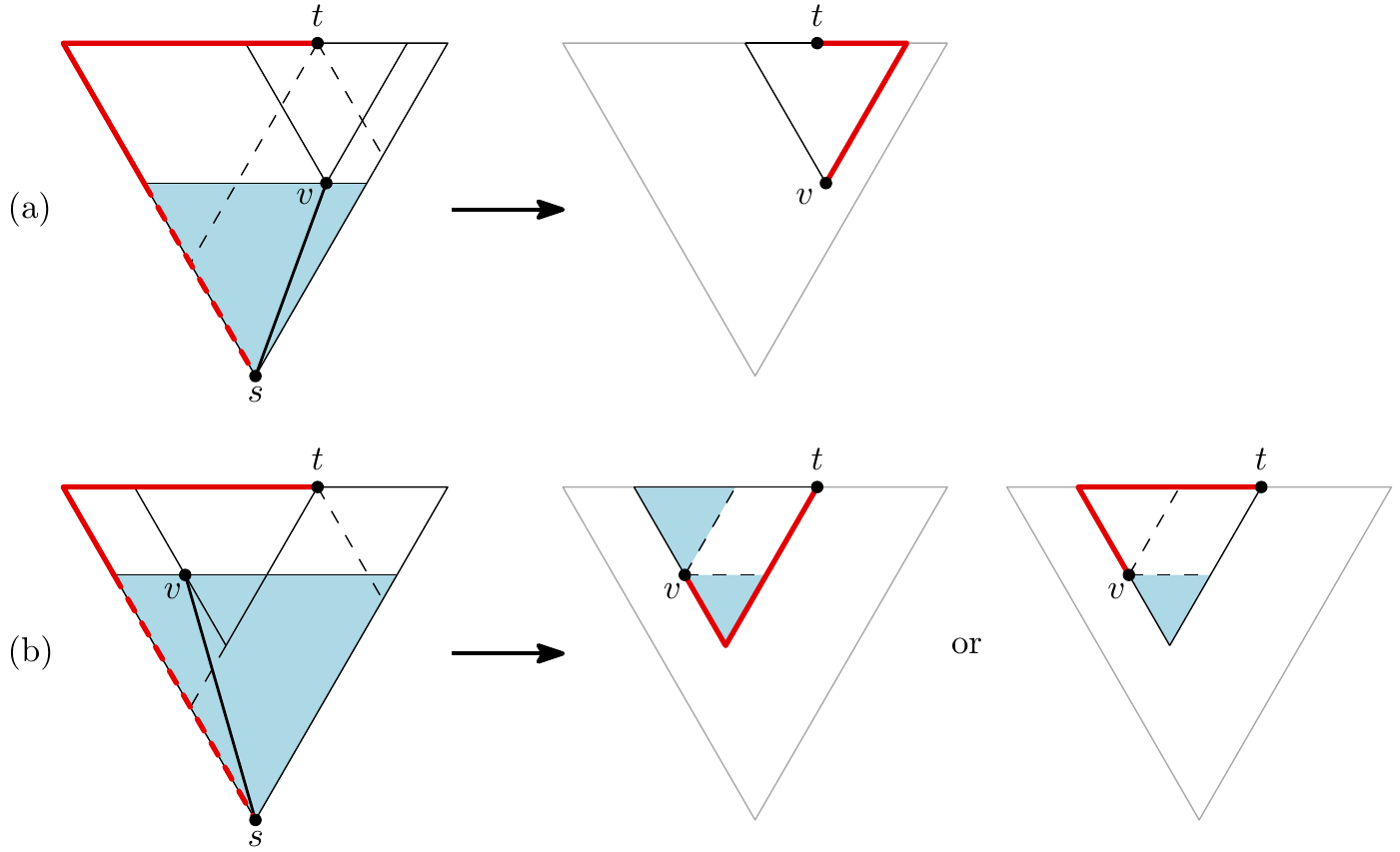}
  \end{center}
  \caption{Routing in case A. (a) $v$ lies in a negative cone of $t$, (b) $v$ lies in a positive cone of $t$. Dashed red lines indicate which parts of the potential are used to pay for the edge.}
  \label{fig:routingStateA}
\end{figure}

\paragraph{Routing algorithm.} While routing, let $s$ denote the current vertex and let $t$ denote the fixed destination (i.e. $t$ corresponds to $w$ in Theorem~\ref{thm:routing}). To be deterministic, $1$-local, and $0$-memory, the routing algorithm needs to determine which edge $(s,v)$ to follow next based only on $s$, $t$, and the neighbours of $s$. We say we are \emph{routing positively} when $t$ is in a positive cone of $s$, and \emph{routing negatively} when $t$ is in a negative cone. (Note the distinction between ``positive routing'' and ``routing positively'': the first describes the conditions \emph{at the start} of the routing process, while the second does so \emph{during} the routing process. In other words, positive routing describes a routing process that starts by routing positively. It is very common for positive routing to include situations where we are routing negatively, see e.g. Figure~\ref{fig:routingStateA}b.)

\begin{figure}[ht]
  \begin{center}
   \includegraphics{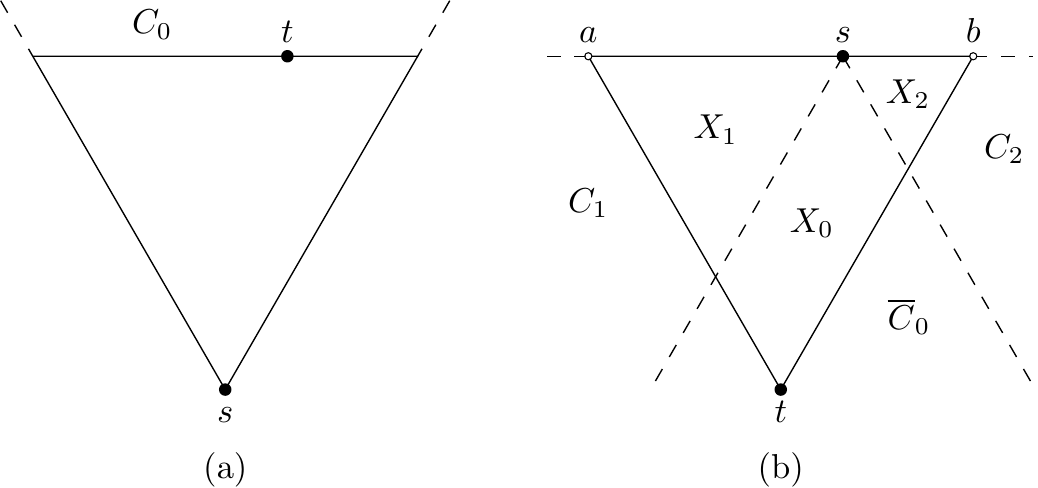}
  \end{center}
  \caption{Routing terminology when (a) routing positively and (b) routing negatively.}
  \label{fig:routingTerminology}
\end{figure}

For ease of description, we assume without loss of generality that $t$ is in cone~$C_0^s$ when routing positively, and in cone~$\overline{C}_0^s$ when routing negatively. When routing positively, \canon{s}{t} intersects only $C_0^s$ among the cones of~$s$. When routing negatively, \canon{t}{s} intersects $\overline{C}_0^s$, as well as the two positive cones $C_1^s$ and $C_2^s$. Let $X_0 = \overline{C}_0^s \cap \canon{t}{s}$, $X_1 = C_1^s \cap \canon{t}{s}$, and $X_2 = C_2^s \cap \canon{t}{s}$. Let $a$ be the corner of~\canon{t}{s} contained in $X_1$ and $b$ the corner of~\canon{t}{s} contained in $X_2$. These definitions are illustrated in Figure~\ref{fig:routingTerminology}.

The routing algorithm will only follow edges~$(s,v)$ where $v$ lies in the canonical triangle of $s$ and $t$.  Routing positively is straightforward since there is exactly one edge $(s,v)$ with $v \in \canon{s}{t}$, by the construction of the \graph. The challenge is to route negatively. When routing negatively, at least one edge $(s,v)$ with $v \in \canon{t}{s}$ exists, since by Theorem~\ref{theo:UnconstrainedSpanningRatio}, $s$ and $t$ are connected by a path inside~\canon{t}{s}. The core of our routing algorithm is how to choose which edge to follow when there is more than one. Intuitively, when routing negatively, our algorithm tries to select an edge that makes measurable progress towards the destination. When no such edge exists, we are forced to take an edge that does not make measurable progress, however we are able to then deduce that certain regions within the canonical triangle are empty. This allows us to bound the total distance traveled while not making measurable progress. We provide a formal description of our routing algorithm below.

Our routing algorithm can be in one of four cases. We call the situation when routing positively case~A, and divide the situation when routing negatively into three further cases: both $X_1$ and $X_2$ are empty (case~B), either $X_1$ or $X_2$ is empty (case~C), or neither is empty (case~D). Since $X_1$ and $X_2$ correspond to positive cones of $s$, each contains the endpoint of at most one edge $(s,v)$. These edges contain a lot of information about the regions $X_1$ and $X_2$. In particular, if there is no edge in the corresponding cone, then the entire cone must be empty. And if there is an edge, but its endpoint lies outside of the region, the region is guaranteed to be empty. This allows our algorithm to \emph{locally} determine if $X_1$ and $X_2$ are empty, and therefore which case we are in.
 
Since we are routing to a destination in a positive cone of the source, our routing algorithm starts in case~A. Routing in this case is straightforward, as there is only one edge $(s,v)$ with $v$ in \canon{t}{s} that we can follow. We now turn our attention to routing in cases~B and~C (it turns out case~D never occurs when routing to a destination in a positive cone of the source; we come back to it when describing negative routing in Section~\ref{sec:negativeRouting}).

In case~B, both $X_1$ and $X_2$ are empty, so there must be edges $(s,v)$ with $v \in X_0$, as $s$ and $t$ are connected by a path in~\canon{t}{s} by Theorem~\ref{theo:UnconstrainedSpanningRatio}. If $\length{as} \ge \length{sb}$, the routing algorithm follows the last edge in clockwise order around $s$; if $\length{as} < \length{sb}$, it follows the first edge. In short, when both sides of~\canon{t}{s} are empty, the routing algorithm favors staying close to the largest empty side of~\canon{t}{s}. Note that $\length{as}$ and $\length{sb}$ can be computed locally from the coordinates of $s$ and $t$.

In case~C, exactly one of $X_1$ or $X_2$ is empty. If there exist edges $(s,v)$ with $v \in X_0$, the routing algorithm will follow one of these, choosing among them in the following way: If $X_1$ is empty, it chooses the last edge in clockwise order around $s$. Else $X_2$ is empty, and it chooses the first edge in clockwise order around $s$. In short, the routing algorithm favors staying close to the empty side of~\canon{t}{s}. If no edges $(s,v)$ with $v \in X_0$ exist, the routing algorithm follows the single edge $(s,v)$ with $v$ in $X_1$ or $X_2$.

\paragraph{Upper bound.} The proof of the upper bound uses a potential function~$\phi$, defined as follows for each of the cases~A, B, and~C. For the potential in case~C, $x \in \{a,b\}$ is the corner contained in the non-empty one of the two areas $X_1$ and $X_2$.

\begin{center}
\begin{tabular}{cl}
Case A:&$\phi = \length{sa} + \max (\length{at},\length{tb})$\\
Case B:&$\phi = \length{ta} + \min (\length{as},\length{sb})$\\
Case C:&$\phi = \length{ta} + \length{sx}$
\end{tabular}
\end{center}

\vspace{-1em}

\begin{figure}[ht]
\begin{center}
\includegraphics{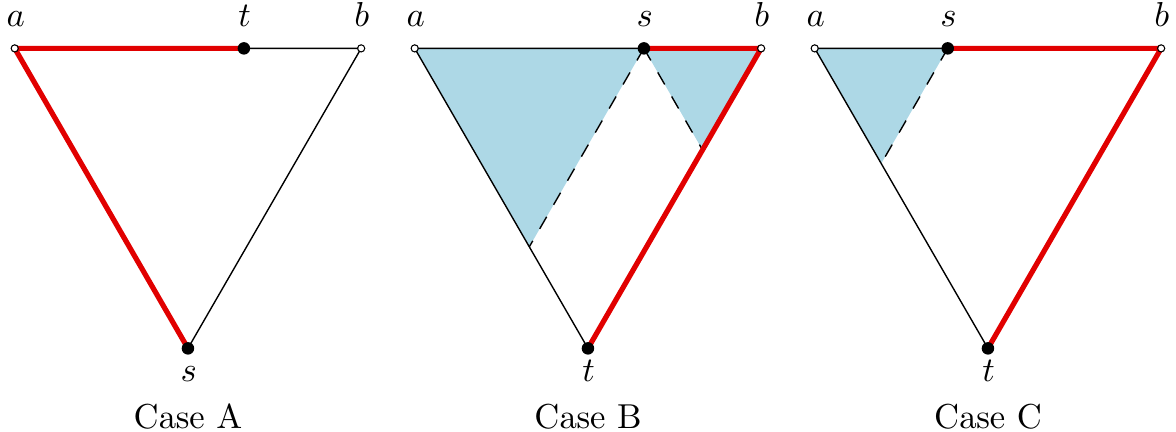}
\end{center}
\caption{The potential $\phi$ in each case. Thick red lines designate potential and light blue designates empty areas.}
\label{fig:potentialFunction}
\end{figure}

This definition is illustrated in Figure~\ref{fig:potentialFunction}. We will refer to the first term of~$\phi$ (i.e., $\length{sa}$ in case~A, $\length{ta}$ in cases~B, and C) as the \emph{vertical part} of~$\phi$ and to the rest as the \emph{horizontal part}. Note that since all sides of the canonical triangle have equal length, $a$ and $b$ are interchangeable in the vertical part. The proof makes extensive use of the following observation about equilateral triangles:

\begin{observation} \label{obs:simpleFact}
In an equilateral triangle, the diameter (the longest distance defined by any two points in the triangle) is equal to the side length.
\end{observation}

Our aim is to prove the following claim: for any routing step, the reduction in $\phi$ is at least as large as the length of the edge followed. This allows us to `pay' for each edge with the difference in potential, thereby bounding the total length of the path by the initial potential. We do this by case analysis of the possible routing steps.

\paragraph{Case~A.} For a routing step starting in case~A, $v$ can be in a negative or a positive cone of~$t$. The first situation leads to case~A again. The second leads to case~B or C, since the area of~\canon{s}{t} between $s$ and $v$ must be empty by construction of the \graph. These situations are illustrated in Figure~\ref{fig:routingStateA}.

If we remain in case~A after following edge $(s, v)$, the reduction of the vertical part of~$\phi$ (dashed in Figure~\ref{fig:routingStateA}a) is at least as large as $\length{sv}$ by Observation~\ref{obs:simpleFact}. Therefore we can use it to pay for this step. Since \canon{v}{t} is contained in \canon{s}{t}, both $\length{at}$ and $\length{bt}$ decrease. Thus the horizontal part of~$\phi$ decreases too, as it is the maximum of the two. Hence the claim holds for this situation.

For the situation ending in case~C (the second illustration after the arrow in Figure~\ref{fig:routingStateA}b), we again use the reduction of the vertical part of~$\phi$ to pay for the step. The rest of the vertical part precisely covers the new horizontal part. Since \canon{t}{v} is contained in \canon{s}{t}, the new vertical part is a portion of either $ta$ or $tb$. This can be covered by the current horizontal part, as it is the maximum of $\length{ta}$ and $\length{tb}$. Thus the claim holds for this situation as well. Finally, for the situation ending in case~B, the final value of $\phi$ is at most that of the situation ending in case~C, so again the claim holds.

\begin{figure}[ht]
  \begin{center}
   \includegraphics{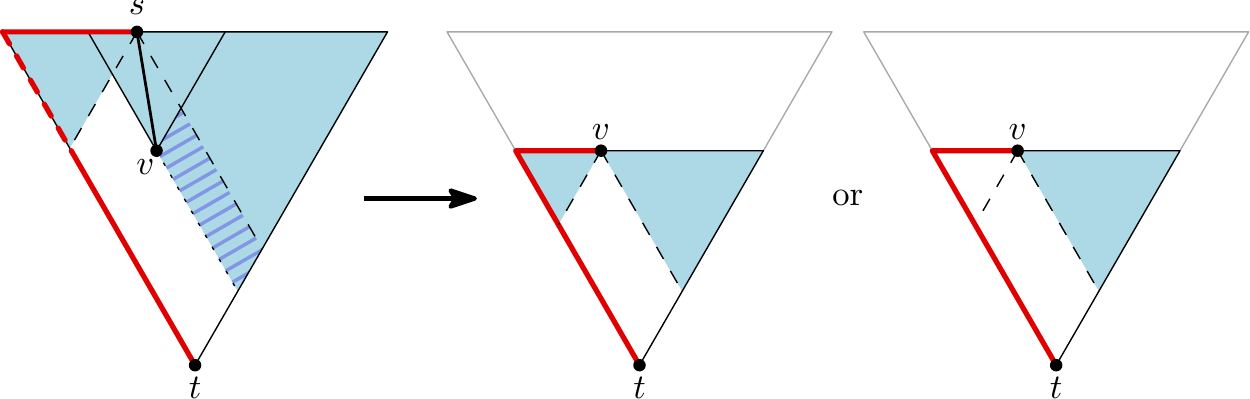}
  \end{center}
  \caption{Routing in case B.}
  \label{fig:routingStateB}
\end{figure}

\paragraph{Case~B.} A routing step starting in case~B (illustrated in Figure~\ref{fig:routingStateB}) cannot lead to case~A, as the step stays within~\canon{t}{s}. We first show that it always results in Case~B or C, meaning that at least one of $X_1$ or $X_2$ is empty again. The algorithm follows an edge $(s,v)$ with $v \in X_0$. If $s$ is to the left of $t$, it follows the first edge in clockwise order around $s$, otherwise it follows the last one. We consider only the case where $s$ is to the left of $t$, the other case is symmetric. By the construction of the \graph, the existence of the edge $(s,v)$ implies that $T_{vs}$ is empty. It follows that the hatched area in Figure~\ref{fig:routingStateB} is also empty: if not, the topmost point in it would have an edge to~$s$, while coming before $v$ in the clockwise order around~$s$, contradicting the choice of~$v$ by the routing algorithm. Therefore $X_2$ will again be empty, resulting in case~B or C.

By Observation~\ref{obs:simpleFact}, the reduction in the vertical part of~$\phi$ is at least as large as $\length{sv}$. In addition, the horizontal part of $\phi$ can only decrease. If it remains on the same side of the triangle, this follows from the fact that $v$ lies in $X_0$ and \canon{t}{v} is contained in \canon{t}{s}. And the only case where the potential switches sides, is when we end up in case~B again but the other side is shorter than the current one, reducing the potential even further. Hence the claim holds.

\begin{figure}[ht]
  \begin{center}
   \includegraphics{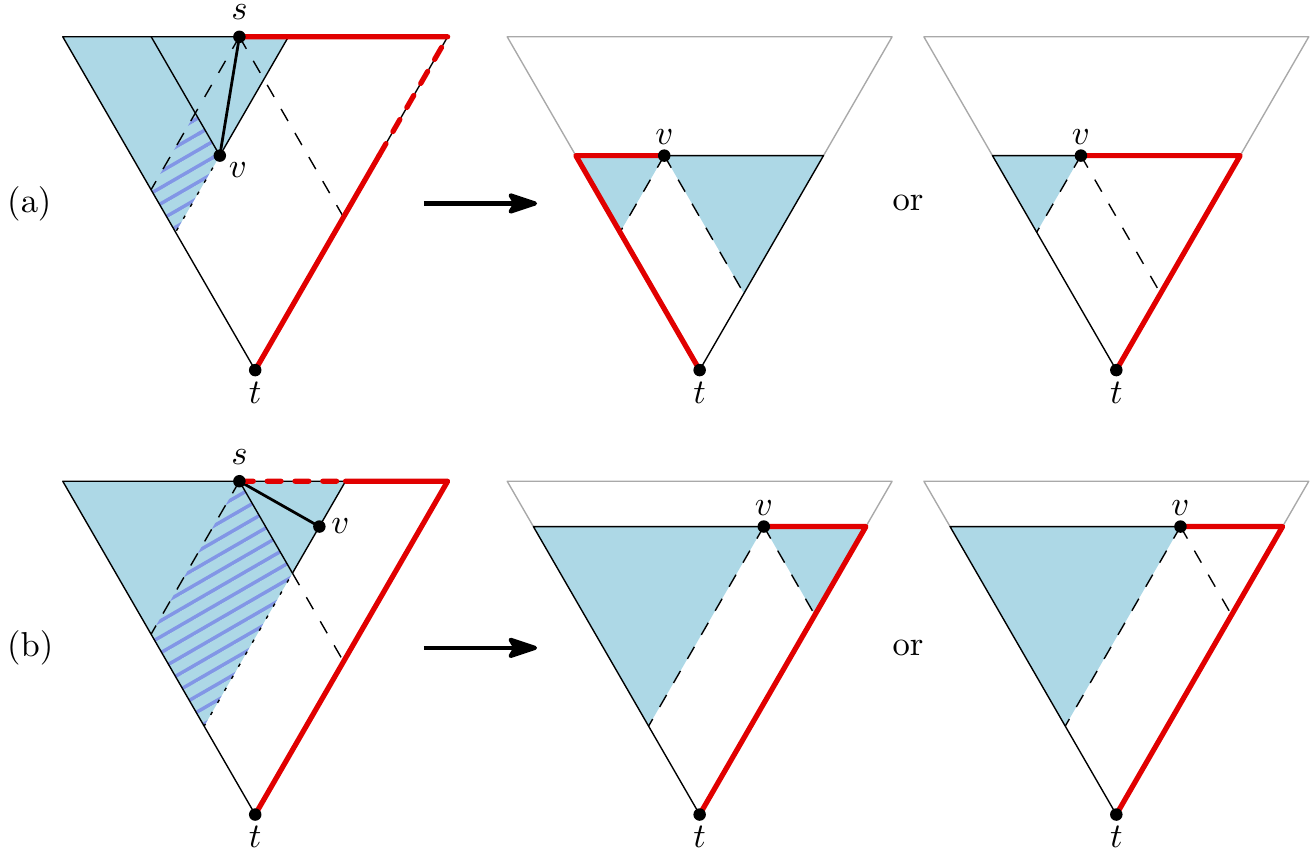}
  \end{center}
  \caption{Routing in case C.}
  \label{fig:routingStateC}
\end{figure}

\paragraph{Case~C.} As in the previous case, a routing step starting in case~C cannot lead to case~A and we show that it cannot lead to case~D, either. There are two situations, depending on whether edges $(s,v)$ with $v \in X_0$ exist. For the situation where such edges do exist (illustrated in Figure~\ref{fig:routingStateC}a), the analysis is exactly the same as for a routing step starting in case~B.

For the situation where edges $(s,v)$ with $v \in X_0$ do not exist, the start of the step is illustrated on the left of the arrow in Figure~\ref{fig:routingStateC}b. Again, $T_{sv}$ must be empty by the construction of the \graph, which implies that the hatched area must also be empty: if not, the topmost point in it would have an edge to~$s$, contradicting that edges $(s,v)$ with $v \in X_0$ do not exist. Thus, the routing step can only lead to case~B or C. Looking at the potential, the vertical part can only decrease, and by Observation~\ref{obs:simpleFact}, the reduction of the horizontal part of~$\phi$ is at least as large as $\length{sv}$. Thus we can pay for this step as well and the claim holds in both situations.

\begin{lemma}\textbf{\emph{(Upper bound for positive routing)}} \label{lem:posub}
 Let $u$ and $w$ be two vertices, with $w$ in a positive cone of $u$. Let $m$ be the midpoint of the side of \canon{u}{w} opposing $u$, and let $\alpha$ be the unsigned angle between $uw$ and $um$. There is a deterministic $1$-local $0$-memory routing algorithm on the \graph for which every path followed has length at most $(\sqrt{3} \cdot \cos \alpha + \sin \alpha) \cdot |u w|$ when routing from $u$ to $w$.
\end{lemma}
\begin{myproof}
 That the algorithm is deterministic, $1$-local, and $0$-memory follows from the description of the algorithm, so we only need to prove the bound on the distance. We showed that for any routing step, the reduction in $\phi$ is at least as large as the length of the edge followed. Since $\phi$ is always non-negative, this implies that no path followed can be longer than the initial value of $\phi$. As all edges have strictly positive length, the routing algorithm must terminate. Since we are routing to a vertex in a positive cone, we start in case~A, with an initial potential of $\length{ua} + \max (\length{aw},\length{wb})$. Taking the side of $T_{uw}$ as the unit of length reduces this to $1 + 1/2 + |wm|$, and using the same analysis as in Lemma~\ref{lem:poslb}, we obtain the desired bound of $(\sqrt{3} \cdot \cos \alpha + \sin \alpha) \cdot |u w|$.
\end{myproof}

\subsection{Negative routing}
\label{sec:negativeRouting}

Next we turn our attention to the case when we are routing to a destination in a negative cone of the source. We start by deriving a lower bound, then present the required extensions to our routing algorithm and finish with the matching upper bound.

\begin{figure}[ht]
  \begin{center}
   \includegraphics{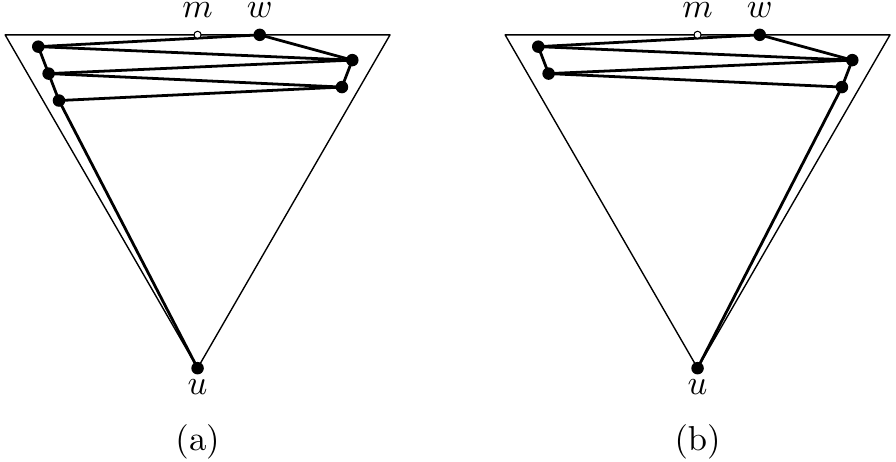}
  \end{center}
  \caption{The lower bound instances for routing to a vertex in a negative cone.}
  \label{fig:lowerBoundInstances}
\end{figure}

\begin{lemma}\textbf{\emph{(Lower bound for negative routing)}} \label{lem:neglb}
 Let $u$ and $w$ be two vertices, with $w$ in a positive cone of $u$. Let $m$ be the midpoint of the side of \canon{u}{w} opposing $u$, and let $\alpha$ be the unsigned angle between $uw$ and $um$. For any deterministic $k$-local $0$-memory routing algorithm, there are instances for which the path followed has length at least $(5/\sqrt{3} \cdot \cos \alpha - \sin \alpha) \cdot |u w|$ when routing from $w$ to $u$.
\end{lemma}
\begin{myproof}
 Consider the two instances in~Figure~\ref{fig:lowerBoundInstances}. Any deterministic $1$-local $0$-memory routing algorithm has information about direct neighbors only. Hence, it cannot distinguish between the two instances when routing out of~$w$. This means that it routes to the same neighbor of $w$ in both instances, and either choice of neighbor leads to a non-optimal route in one of the two instances. The smallest loss occurs when the choice is towards the closest corner of~\canon{u}{w}, for which Figure~\ref{fig:lowerBoundInstances}a is the bad instance. If we let the side of \canon{u}{w} be the unit of length, this gives a lower bound of $(1/2-\length{wm})+1+1 = 5/2-\length{wm}$, since the points in the corners of~\canon{u}{w} can be moved arbitrarily close to the corners while keeping their relative positions. Using that $\length{wm} = \length{uw}\cdot\sin\alpha$ and $\sqrt{3}/2 = \length{um} = \length{uw}\cdot\cos\alpha$, the lower bound reduces to $(5/\sqrt{3} \cdot \cos \alpha - \sin \alpha)\cdot\length{uw}$. By appropriately adding $\Omega(k)$ points close to the corners such that $u$ is not in the $k$-neighborhood of $w$, the lower bound holds for any deterministic $k$-local $0$-memory routing algorithm.
\end{myproof}

\paragraph{Routing algorithm.} The only difference with the routing algorithm we used for positive routing lies in the initial case. Since our destination is in a negative cone, we start in one of the negative cases. This time, besides cases~B and C, where both or one of $X_1$ and $X_2$ are empty, we also need case~D, where neither is empty. Recall that in the previous section, we showed that a routing step starting in case~A, B, or C can never result in case~D. Thus, if the routing process starts in case~D, it never returns there once it enters case~A, B, or C.

In case~D, the routing algorithm first tries to follow an edge $(s,v)$ with $v \in X_0$. If several such edges exist, an arbitrary one of these is followed. If no such edge exists, the routing algorithm follows the single edge $(s,v)$ with $v$ in the smaller of $X_1$ and $X_2$. In short, the routing algorithm favors moving towards the closest corner of~\canon{t}{s} when it is not able to move towards~$t$. Note that, in the instances of Figure~\ref{fig:lowerBoundInstances}, this choice ensures that the first routing step incurs the smallest loss in the worst case, making it possible to meet the lower bound of Lemma~\ref{lem:neglb}. We now show that our algorithm achieves this lower bound in all cases.

\paragraph{Upper bound.} The potential in case~D is given below. It mirrors the lower bound path, in that it allows walking towards the closest corner, crossing the triangle, then walking down to $t$. This is the highest potential among the four cases.

\begin{center}
\begin{tabular}{cl}
Case D:&$\phi = \length{ta} + \length{ab} + \min (\length{as},\length{sb})$
\end{tabular}
\end{center}

\begin{figure}[ht]
\begin{center}
\includegraphics{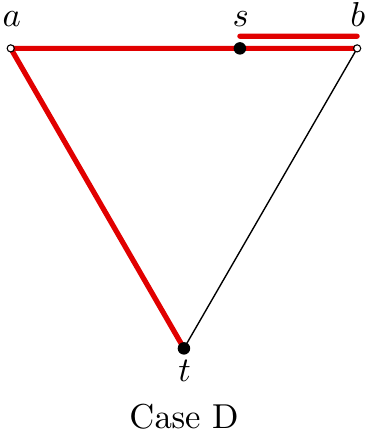}
\end{center}
\caption{The potential $\phi$ in case~D.}
\label{fig:potentialFunctionNeg}
\end{figure}

As before, we want to show that for any routing step, the reduction in $\phi$ is at least as large as the length of the edge followed. Since we already did this for states~A, B, and C, all that is left is to prove it for case~D.

\begin{figure}[ht]
  \begin{center}
   \includegraphics{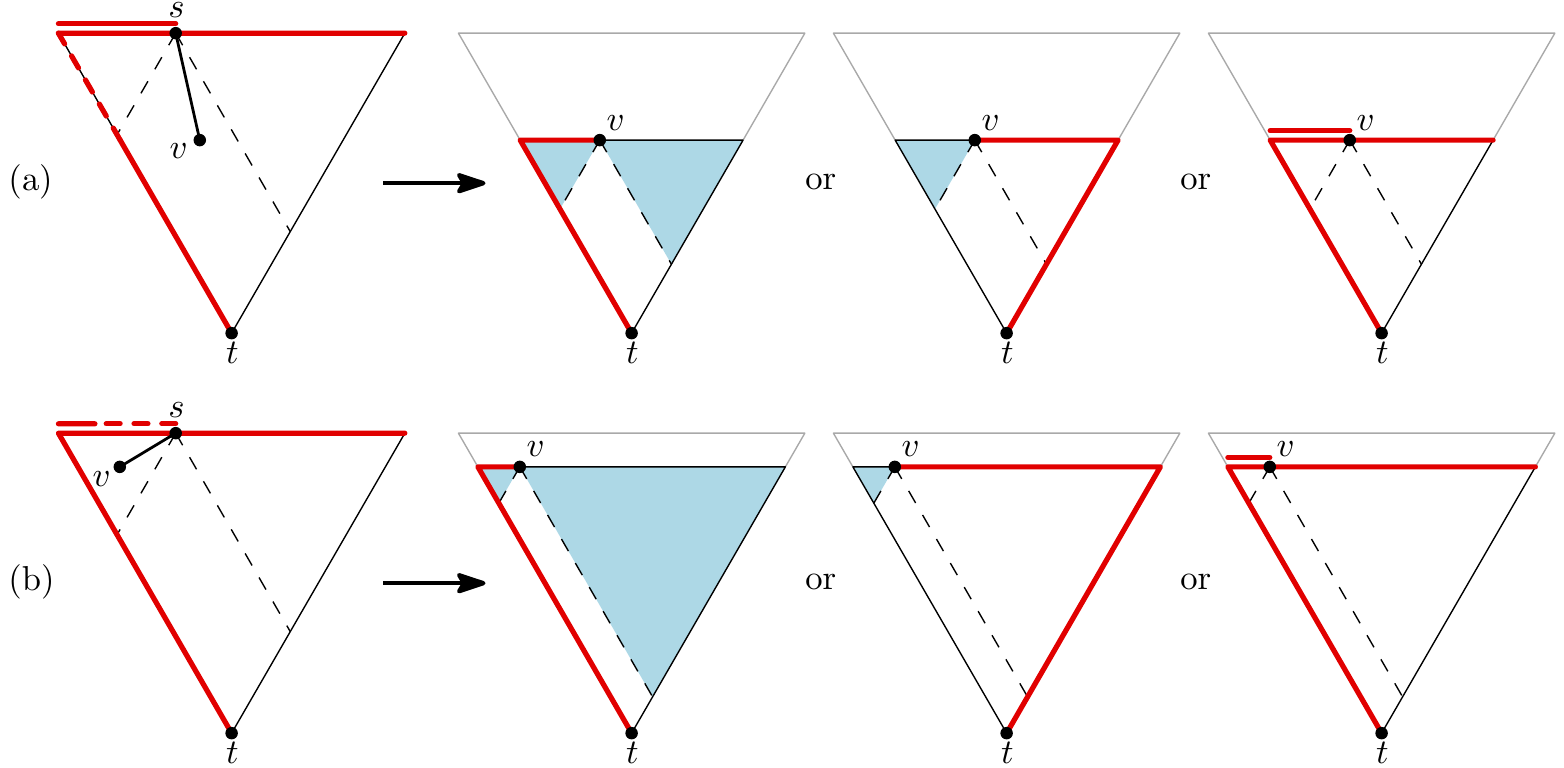}
  \end{center}
  \caption{Routing in case D. The endpoint $v$ of the edge followed lies in $X_0$ (a), or the smaller of $X_1$ and $X_2$ (b).}
  \label{fig:routingStateD}
\end{figure}

\paragraph{Case~D.} A routing step starting in case~D cannot lead to case~A, as the step stays within~\canon{t}{s}, but it may lead to case~B, C, or D. There are two situations, depending on whether edges $(s,v)$ with $v \in X_0$ exist or not. These are illustrated in Figure~\ref{fig:routingStateD}.

In the first situation, where we follow an edge $(s, v)$ with $v \in X_0$, the reduction of the vertical part of~$\phi$ is at least as large as $\length{sv}$ by Observation~\ref{obs:simpleFact}. The horizontal part of~$\phi$ can only decrease, as \canon{t}{v} is fully contained in \canon{t}{s} and $v$ lies in $X_0$. In the second situation, where the endpoint of our edge lies in the smaller of $X_1$ and $X_2$, these roles switch, with the reduction of the horizontal part of~$\phi$ being at least as large as $\length{sv}$ and the vertical part of~$\phi$ only decreasing. In both situations, the statement is proven.

\begin{lemma}\textbf{\emph{(Upper bound for negative routing)}} \label{lem:negub}
 Let $u$ and $w$ be two vertices, with $w$ in a positive cone of $u$. Let $m$ be the midpoint of the side of \canon{u}{w} opposing $u$, and let $\alpha$ be the unsigned angle between $uw$ and $um$. There is a deterministic $1$-local $0$-memory routing algorithm on the \graph for which every path followed has length at most $(5/\sqrt{3} \cdot \cos \alpha - \sin \alpha) \cdot |u w|$ when routing from $w$ to $u$.
\end{lemma}
\begin{myproof}
 Since the choices that the routing algorithm makes are completely determined by the neighbours of $s$ and the location of $s$ and $t$, the algorithm is indeed deterministic, $1$-local, and $0$-memory. To bound the length of the resulting path, we again showed that for any routing step, the reduction in $\phi$ is at least as large as the length of the edge followed. As in the proof of Lemma~\ref{lem:posub}, this implies that the routing algorithm terminates and that the total length of the path followed is bounded by the initial value of $\phi$. Since our destination lies in a negative cone, we start in one of the cases~B, C, or D. Of these three cases, case~D has the largest initial potential of $\length{ta} + \length{ab} + \min (\length{as},\length{sb})$. Taking the side of $T_{uw}$ as the unit of length reduces this to $1 + 1 + 1/2 - \length{wm} = 5/2 - \length{wm}$, and using the same analysis as in Lemma~\ref{lem:neglb}, we obtain the desired bound of $(5/\sqrt{3} \cdot \cos \alpha - \sin \alpha) \cdot |u w|$.
\end{myproof}

\noindent As Theorem~\ref{thm:routing} follows from Lemmas~\ref{lem:poslb}, \ref{lem:posub}, \ref{lem:neglb}, and \ref{lem:negub}, this concludes our proof.

\section{A stateful algorithm}
\label{sec:stateful}

Next we present a slightly different routing algorithm from the one in the previous section. The main difference between the two algorithms is that this one maintains one piece of information as state, making it $O(1)$-memory instead of $0$-memory. The information that is stored is a \emph{preferred side}, and it is either nil, $X_1$, or $X_2$. Intuitively, the new algorithm follows the original algorithm until it is routing negatively and determines that either $X_1$ or $X_2$ is empty. At that point, the algorithm sets the empty side as the preferred side and picks the rest of the edges in such a way that the preferred side remains empty. Thus, the algorithm maintains as invariant that if the preferred side is set (not nil), that region is empty. Furthermore, once the preferred side is set, it stays fixed until the algorithm reaches the destination. This algorithm simplifies the cases a little, but more importantly, it allows the algorithm to check far fewer edges while routing. This is crucial, as the new algorithm forms the basis for routing algorithms on versions of the \graph with some edges removed to bound the maximum degree, described in the next section.

We now present the details of this stateful version of the routing algorithm. Recall that we are trying to find a path from a current vertex $s$ to a destination vertex $t$. For ease of description, we again assume without loss of generality that $t$ lies in $\c{0}$ or $\nc{0}$ of $s$. If $t$ lies in $\nc{0}$, the cones around $s$ split $T_{ts}$ into three regions $X_0$, $X_1$, and $X_2$, as in Figure~\ref{fig:routingTerminology}. For brevity, we use ``an edge in $X_0$'' to denote an edge incident to $s$ with the other endpoint in $X_0$. The cases are as follows:

\begin{shortitemize}
 \item If $t$ lies in a positive cone of $s$, we are in case~\sA.
 \item If $t$ lies in a negative cone of $s$ and no preferred side has been set yet, we are in case~\sB.
 \item If $t$ lies in a negative cone of $s$ and a preferred side has been set, we are in case~\sC.
\end{shortitemize}

These cases are closely related to the cases in the stateless algorithm. Cases~\sA and \sB correspond to cases~A and D, respectively, while case~\sC merges cases~B and C from the original algorithm into a single case, where only one side's emptiness is tracked. This is reflected in the routing strategy for each case:

\begin{shortitemize}
 \item In case~\sA, follow the unique edge $(s, v)$ in the positive cone containing $t$. If $t$ lies in a negative cone of $v$, set the preferred side to the region ($X_1$ or $X_2$ of $v$) that is contained in $T_{sv}$, as this is now known to be empty (see Figure~\ref{fig:routingStateA}b).
 \item In case~\sB, if there are edges in $X_0$, follow an arbitrary one. Otherwise, if there is an edge in the smaller of $X_1$ and $X_2$, follow that edge. Otherwise, follow the edge in the larger of $X_1$ and $X_2$ and set the other as the preferred side. By Theorem~\ref{theo:UnconstrainedSpanningRatio}, at least one of these edges must exist.
 \item In case~\sC, if there are edges in $X_0$, follow the one closest to the preferred side in cyclic order around $s$. Otherwise, follow the edge in the positive cone that is not on the preferred side. Again, at least one of these edges must exist.
\end{shortitemize}

The proof in Section~\ref{sec:routing} can be adapted to show that this routing algorithm achieves the same upper bounds. In short, the proof is simplified to only use a potential as defined for cases~A, C, and D, and only a subset of the illustrations in Figures~\ref{fig:routingStateA}, \ref{fig:routingStateC}, and \ref{fig:routingStateD} are relevant. We omit the repetitive details.

\section{Bounding the maximum degree}
\label{sec:boundeddeg}

Each vertex in the \hts has at most one incident edge in each positive cone, but it can have an unbounded number of incident edges in its negative cones. In this section, we describe two transformations that allow us to bound the total degree of each vertex. The transformations are adapted from Bonichon~\etal~\cite{BGHP10}.

\begin{figure}[ht]
 \centering
 \includegraphics{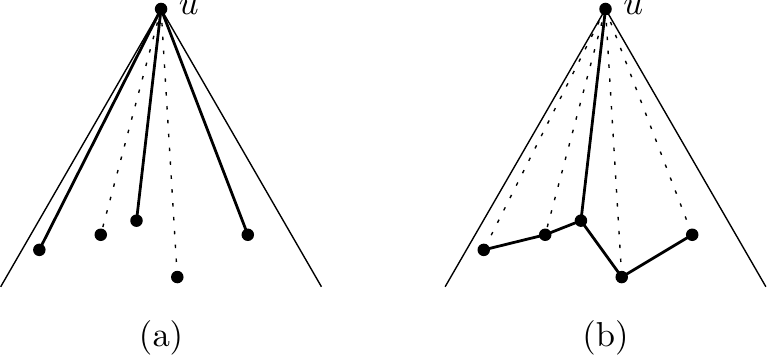}
 \caption{The construction for \dtw (a) and \dn (b). Solid edges are kept, while dotted edges are discarded if no other vertex wants to keep them.}
 \label{fig:boundeddegree}
\end{figure}

The first transformation discards all edges in each negative cone, except for three: the first and last edges in clockwise order around the vertex and the edge to the ``closest'' vertex, meaning the vertex whose projection on the bisector of the cone is closest (see Figure~\ref{fig:boundeddegree}a). This results in a subgraph with maximum degree 12, which we call \dtw.

To reduce the degree even further, we note that since the \hts is internally triangulated, consecutive neighbours of $u$ within a negative cone are connected by edges. We call the path formed by these edges the \emph{canonical path}. Instead of keeping three edges per negative cone, we now keep only the edge to the closest vertex, but force the edges of the canonical path to be kept as well (see Figure~\ref{fig:boundeddegree}b). We call the resulting graph \dn. Bonichon~\etal~\cite{BGHP10} showed that all edges on the canonical path are either first or last in a negative cone, making \dn a subgraph of \dtw. Note that since the \hts is planar, both subgraphs are planar as well. They also proved that \dn is a 3-spanner of the \hts with maximum degree 9. Since the \hts is a 2-spanner and \dn is a subgraph of \dtw, this shows that both \dn and \dtw are 6-spanners of the complete Euclidean graph. We give an adapted version of the proof of the spanning ratio of \dn below.

\begin{theorem}
 \label{thm:3spanner}
 \dn is a 3-spanner of the \hts.
\end{theorem}
\begin{myproof}
 Consider an edge $(s, v)$ in the \hts and assume, without loss of generality, that $v$ lies in a negative cone of $s$ (if not, we can swap the roles of $s$ and $v$). Now consider the path between them in \dn{} consisting of the edge from $s$ to the vertex closest to $s$, followed by the edges on the canonical path between the closest vertex and $v$. We will refer to this path as the \emph{approximation path}, and we show that it has length at most $3 \cdot |sv|$.

 \begin{figure}[ht]
  \centering
  \includegraphics{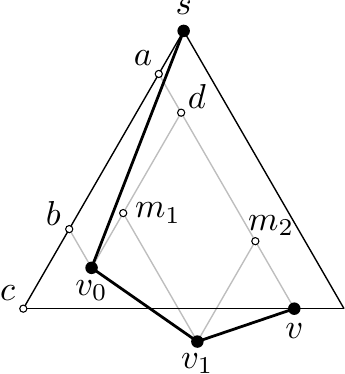}
  \caption{The approximation path.}
  \label{fig:spanningratio}
 \end{figure}

 Let $v_0$ be the closest vertex and let $v_1, \dots, v_k = v$ be the other vertices on the approximation path. We assume without loss of generality that $s$ lies in $\c{0}$ of $v$ and that $v$ lies to the right of $v_0$. We shoot rays parallel to the boundaries of $\c{0}$ from each vertex on the approximation path. Let $m_i$ be the intersection of the right ray of $v_{i-1}$ and the left ray of $v_i$ (see Figure~\ref{fig:spanningratio}). These intersections must exist, as $s$ is the closest vertex in $\c{0}^{v_i}$, for each $v_i$. Let $a$ and $b$ be the intersections of the left boundary of $\nc{0}^s$ with the left rays of $v$ and $v_0$, respectively, and let $c$ be the intersection of this left boundary with the horizontal line through $v$. Finally, let $d$ be the intersection of the right ray of $v_0$ and the left ray of $v$. We can bound the length of the approximation path as follows:
 \begin{eqnarray*}
  & & |sv_0| + \sum_{i=1}^k |v_{i-1}v_i| \\
  & \leq & |sb| + |bv_0| + \sum_{i=1}^k |v_{i-1}m_i| + \sum_{i=1}^k |m_iv_i| \\
  & = & |sb| + |bv_0| + |ab| + |dv| \textmd{~~~~\{by projection\}}\\
  & = & |sb| + |ab| + |av| \\
  & \leq & |sc| + 2 \cdot |cv|
 \end{eqnarray*}

 The last inequality follows from the fact that $v_0$ is the closest vertex to $s$. Let $\alpha$ be $\angle csv$. Some basic trigonometry gives us that $|sc| = \frac{2}{\sqrt{3}} \cdot \sin\left(\alpha + \frac{\pi}{3}\right) \cdot |sv|$ and $|cv| = \frac{2}{\sqrt{3}} \cdot \sin \alpha \cdot |sv|$. Thus the approximation path is at most $\frac{2}{\sqrt{3}} \cdot \left(\sin\left(\alpha + \frac{\pi}{3}\right) + 2 \cdot \sin \alpha \right)$ times as long as $(s, v)$. Since this function is increasing in $[0, \frac{\pi}{3}]$, the maximum is achieved for $\alpha = \pi/3$, where it is 3. Therefore every edge of the \hts can be approximated by a path that is at most 3 times as long and the theorem follows.
\end{myproof}

Note that the part of the approximation path that lies on the canonical path has length at most $2 \cdot |cv| = \frac{4}{\sqrt{3}} \cdot \sin \alpha \cdot |sv|$. This function is also increasing in $[0, \frac{\pi}{3}]$ and its maximal value is 2, so the total length of this part is at most $2 \cdot |sv|$.

\subsection{Routing in $\boldsymbol{\dtw}$}
\label{sec:dtw}

The stateful algorithm in Section~\ref{sec:stateful} constructs a path between two vertices in the \hts. We cannot directly follow this path in \dtw, as some of the edges may have been removed. Hence, we need to find a new path in \dtw that approximates the path in the \hts, taking the missing edges into account. This often amounts to following the approximation path for edges that are in the path in the \hts, but were removed to create \dtw. In addition, some of the information the algorithm uses to decide which edge to follow relies on the presence or absence of edges in the \hts. Since the absence of these edges in \dtw does not tell us whether or not they were present in the \hts, we need to find a new way to make these decisions.

First, note that the only information we need to determine in which of the three cases we are, are the coordinates of $s$ and $t$ and whether the preferred side has been set or not. Therefore we can still make this distinction in \dtw. The following five headlines refer to steps of the stateful algorithm on the \hts, and the text after a headline describes how to simulate that step in \dtw. We discuss modifications for \dn in Section~\ref{sec:dn}.

\paragraph{Follow an edge $\boldsymbol{(s, v)}$ in a positive cone $\boldsymbol{C}$.} If the edge of the \hts is still present in \dtw, we simply follow it. If it is not, the edge was removed because $s$ is on the canonical path of $v$ and it is not the closest, first or last vertex on the path. Since \dtw is a supergraph of \dn, we know that all of the edges of the canonical path are kept and every vertex on the path originally had an edge to $v$ in $C$. Therefore it suffices to traverse the canonical path in one direction until we reach a vertex with an edge in $C$, and follow this edge. Since the edges connecting $v$ to the first and last vertices on the path are always kept, the edge we find in this way must lead to~$v$. Note that the edges of the canonical path are easy to identify, as they are the closest edges to $C$ in cyclic order around $s$ (one on either side of $C$).

This method is guaranteed to reach $v$, but we want to find a \emph{competitive} path to $v$. Therefore we use exponential search along the canonical path: we start by following the shorter of the two edges of the canonical path incident to $s$. If the endpoint of this edge does not have an edge in $C$, we return to $s$ and travel twice the length of the first edge in the other direction. We keep returning to $s$ and doubling the maximum travel distance until we find a vertex $x$ that does have an edge in $C$. If $x$ is not the closest to $v$, by the triangle inequality, following its edge to $v$ is shorter than continuing our search until we reach the closest and following its edge. So for the purpose of bounding the distance travelled, we can assume that $x$ is closest to $v$. Let $d$ be the distance between $s$ and $x$ along the canonical path. By using exponential search to find $x$, we travel at most 9 times this distance~\cite{MR1241311} and afterwards we follow $(x, v)$. From the proof of Theorem~\ref{thm:3spanner}, we know that $d \leq 2 \cdot |sv|$ and $d + |xv| \leq 3 \cdot |sv|$. Thus the total length of our path is at most $9 \cdot d + |xv| = 8 \cdot d + (d + |xv|) \leq 16 \cdot |sv| + 3 \cdot |sv| = 19 \cdot |sv|$.

\begin{figure}[ht]
 \centering
 \includegraphics{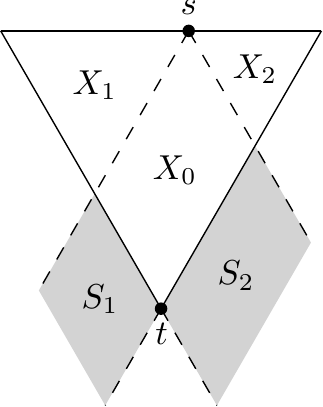}
 \caption{Possible regions for the first and last vertex.}
 \label{fig:checkx0}
\end{figure}

\paragraph{Determine if there are edges in $\boldsymbol{X_0}$.} In the regular \hts we can look at all our neighbours and see if any of them lie in $X_0$. However, in \dtw, these edges may have been removed. Fortunately, we can still determine if they existed in the original \hts. To do this, we look at the vertices of the canonical path in this cone that are first and last in clockwise order around $s$. If these vertices do not exist, $s$ did not have any incoming edges in this cone, so there can be no edges in $X_0$. If the first and last are the same vertex, this was the only incoming edge to $s$ from this cone, so we simply check if its endpoint lies in $X_0$. The interesting case is when the first and last exist and are distinct. If either of them lies in $X_0$, we have our answer, so assume that both lie outside of $X_0$. Since they were connected to $s$, they cannot have $t$ in their positive cone, so they must lie in one of two regions, which we call $S_1$ and $S_2$ (see Figure~\ref{fig:checkx0}).

If both the first and last lie in $S_2$, there can be no edge in $X_0$, since any vertex of the canonical path in $X_0$ either lies in cone $\c{0}$ of the last vertex, or would come after the last vertex in clockwise order around $s$. Both yield a contradiction. If both lie in $S_1$, a similar argument using the first vertex applies.

On the other hand, if the first lies in $S_2$ and the last in $S_1$, both $X_1$ and $X_2$ have to be empty, since both vertices are connected to $s$. Now we are in one of two cases: either $X_0$ is also empty, or it is not. If there are no vertices in $X_0$ (different from~$t$ and $s$), $t$ must have had an edge to $s$. On the other hand, if there are other vertices in $X_0$, the topmost of these vertices must have had an edge to $s$. In either case, there must have been an edge in $X_0$. This shows that we can check whether there was an edge in $X_0$ in the \hts using only the coordinates of the first and last vertex.

\paragraph{Follow an arbitrary edge in $\boldsymbol{X_0}$.} If the \hts has edges in $X_0$, we simulate following an arbitrary one of these by first following the edge to the closest vertex in the negative cone. If this vertex is in $X_0$, we are done. Otherwise, we follow the canonical path in the direction of $X_0$ and stop once we are inside. This traverses exactly the approximation path of the edge, and hence travels a distance of at most 3 times the length of the edge.

\begin{figure}[ht]
 \centering
 \includegraphics{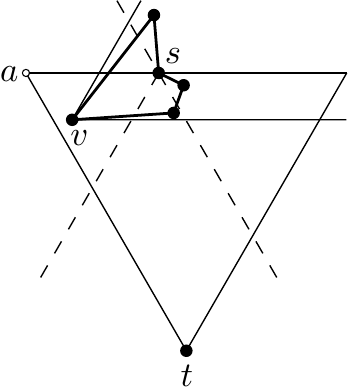}
 \caption{An example where $s$ had an edge in $X_1$ in the \hts, which was removed during the construction of \dtw.}
 \label{fig:checkx1}
\end{figure}

\paragraph{Determine if there is an edge in $\boldsymbol{X_1}$ or $\boldsymbol{X_2}$.} Since these regions are symmetric, we will consider only the case for $X_1$. Since $X_1$ is contained in a positive cone of $s$, it contains at most one edge incident to $s$. If the edge is present in \dtw, we can simply test whether the other endpoint lies in $X_1$. However, if $s$ does not have a neighbour in this cone (see Figure~\ref{fig:checkx1}), we need to find out whether it used to have one in the original \hts and if so, whether it was in $X_1$. Since this step is only needed in case~\sB after we determine that there are no edges in $X_0$, we can use this information to guide our search. Specifically, we know that if we find an edge, we should follow it.

Therefore we simply attempt to follow the edge in this cone, using the exponential search method for following an edge in a positive cone described earlier. Let $x$ be the first vertex we encounter that still has an edge $(x,w)$ in $C_1$. If in the \hts, $s$ had an edge $(s, v)$ in $X_1$, then we know (from the arguments presented earlier for following an edge in a positive cone) that $w$ is $v$. As such, $w$ must lie in $X_1$. We also know (from the proof of Theorem~\ref{thm:3spanner}) that the distance along the canonical path from $s$ to $x$ is at most $2 \cdot |sv|$, which is bounded by $2 \cdot |as|$ since $v$ lies in $X_1$. In this case, we follow the edge from $x$ to $v$. Conversely, if we do not find any vertex with an edge in $C_1$ within a distance of $2 \cdot |as|$ from $s$, or we do, but the endpoint of the edge ($w$) does not lie in $X_1$, then we can return to $s$ and conclude that it did not have an edge in $X_1$ in the \hts and therefore $X_1$ must be empty.

If there was an edge in $X_1$, we travelled the same distance as if we were simply following the edge: at most $19 \cdot |sv|$. If we return to $s$ unsuccessfully, we travelled at most $20 \cdot |as|$: 9 times $2 \cdot |as|$ during the exponential search and $2 \cdot |as|$ to return to $s$.

\paragraph{Follow the edge in $\boldsymbol{X_0}$ closest to the preferred side in clockwise order.} To follow this edge, we first follow the edge to the closest vertex. If this lands us in $X_0$, we then follow the canonical path towards the preferred side and stop at the last vertex on the canonical path that is in $X_0$. If the closest is not in $X_0$, we follow the canonical path towards $X_0$ and stop at the first or last vertex in $X_0$, depending on which side of $X_0$ we started on. This follows the approximation path of the edge, so the distance travelled is at most 3 times the length of the edge.

\paragraph{Routing ratio.} This shows that we can simulate the stateful routing algorithm on \dtw. As state in the message, we need to store not only the preferred side, but also information for the exponential search, including distance travelled. The exact routing ratios are as follows.

\begin{theorem}
 Let $u$ and $w$ be two vertices, with $w$ in a positive cone of $u$. There exists a deterministic $1$-local $O(1)$-memory routing algorithm on \dtw with routing ratio
 \renewcommand{\labelenumi}{{\rm\roman{enumi})}}
 \begin{enumerate}
  \item $19 \cdot 2 = 38$ when routing from $u$ to $w$,
  \item $19 \cdot 5/\sqrt{3} = 54.848\dots$ when routing from $w$ to $u$.
 \end{enumerate}
\end{theorem}
\begin{myproof}
 As shown above, we can simulate every edge followed by the algorithm by travelling at most 19 times the length of the edge. The only additional cost is incurred in case~\sB, when we try to follow an edge in the smaller of $X_1$ and $X_2$, but this edge does not exist. In this case, we travel an additional $20 \cdot |as|$, where $a$ is the corner closest to $s$. Fortunately, this can happen at most once during the execution of the algorithm, as it prompts the transition to case~\sC, after which the algorithm never returns to case~\sB. Looking at the proof for the upper bound in Section~\ref{sec:routing} (specifically, the second case in Figure~\ref{fig:routingStateD}b), we observe that in the transition from case~$D$ to $C$, there is $2 \cdot |as|$ of unused potential. Since we are trying to show a routing ratio of 19 times the original, we can charge the additional $20 \cdot |as|$ to the $38 \cdot |as|$ of unused potential.
\end{myproof}

\subsection{Routing in $\boldsymbol{\dn}$}
\label{sec:dn}

In this subsection, we explain how to modify the previously described simulation strategies so that they work for \dn, where the first and last edges are not guaranteed to be present. We discuss only those steps that rely on the presence of these edges. To route successfully in this setting, we need to change our model slightly. We now let every vertex store a constant amount of information in addition to the information about its neighbours.

\paragraph{Follow an edge $\boldsymbol{(s, v)}$ in a positive cone.} Because the first and last edges are not always kept, we cannot guarantee that the first vertex we reach with an edge in this positive cone is still part of the same canonical path. This means that the edge could connect to some arbitrary vertex, far away from $v$. Therefore our original exponential search solution does not work. Instead, we store one bit of information at $s$ (per positive cone), namely in which direction we have to follow the canonical path to reach the closest vertex to $v$. Knowing this, we just follow the canonical path in the indicated direction until we reach a vertex with an edge in this positive cone. This vertex must be the closest, so it gives us precisely the approximation path and therefore we travel at most $3 \cdot |sv|$.

\paragraph{Determine if there are edges in $\boldsymbol{X_0}$.} In \dtw, this test was based on the coordinates of the endpoints of the first and last edge. Since these might be missing in \dn, we store the coordinates of these vertices at $s$. This allows us to perform the check without increasing the distance travelled.

\paragraph{Determine if there is an edge in $\boldsymbol{X_1}$ or $\boldsymbol{X_2}$.} As in the positive routing simulation, we now know where to go to find the closest. Therefore we simply follow the canonical path in this direction from $s$ and stop when we reach a vertex with an edge in the correct positive cone, or when we have travelled $2 \cdot |as|$. If there is an edge, we follow exactly the approximation path, giving us 3 times the length of the edge. If there is no edge, we travel $2 \cdot |as|$ back and forth, for a total of $4 \cdot |as|$.

\paragraph{Routing ratio.} Since the other simulation strategies do not rely on the presence of the first or last edges, we can now analyze the routing ratio obtained on \dn.

\begin{theorem}
 Let $u$ and $w$ be two vertices, with $w$ in a positive cone of $u$. By storing $O(1)$ additional information at each vertex, there exists a deterministic $1$-local $O(1)$-memory routing algorithm on \dn and \dtw with routing ratio
 \renewcommand{\labelenumi}{{\rm\roman{enumi})}}
 \begin{enumerate}
  \item $3 \cdot 2 = 6$ when routing from $u$ to $w$,
  \item $3 \cdot 5/\sqrt{3} = 8.660\dots$ when routing from $w$ to $u$.
 \end{enumerate}
\end{theorem}
\begin{myproof}
 The simulation strategy for \dtw followed the approximation path for each edge, except when following an edge in a positive cone. Since our new strategy follows the approximation path there as well, our new routing ratio is only 3 times the one for the \hts. Note that this is still sufficient to charge the additional $4 \cdot |sa|$ travelled to the transition from case~\sB to~\sC, which has $3 \cdot 2 \cdot |as|$ of otherwise unused potential. Since \dn is a subgraph of \dtw, this strategy works on \dtw as well.
\end{myproof}

\section{Conclusions}

We presented a competitive deterministic $1$-local $0$-memory routing algorithm on the \graph. We also presented matching lower bounds on the routing ratio for any deterministic $k$-local $0$-memory algorithm, showing that our algorithm is optimal. Since any triangulation can be embedded as a \graph using Schnyder's embedding~\cite{S90}, this shows that any triangulation has an embedding that admits a competitive routing algorithm. An interesting open problem here is whether this approach can be extended to other theta-graphs. In particular, we recently extended the proof for the spanning ratio of the \graph to theta-graphs with $4k+2$ cones, for integer $k > 0$~\cite{bose2012optimal}. It would be interesting to see if it is possible to find optimal routing algorithms for these graphs as well.

We further extended our routing algorithm to work on versions of the \graph with bounded maximum degree. As far as we know, these are the first competitive routing algorithms on bounded-degree plane graphs. There are several problems here that are still open. For example, while we found a matching lower bound for negative routing in the regular \graph, we do not have one for the version with bounded degree. Can we find this, or is it possible to improve the routing algorithm further? And can we extend the algorithm to the version with maximum degree 6, introduced by Bonichon~\etal~\cite{BGHP10}?

\newpage
\pagestyle{empty}

\bibliographystyle{abbrv}
\bibliography{route}

\end{document}